# Optimal Parameter Design for Power Electronic Converters Using a Probabilistic Learning-Based Stochastic Surrogate Model

Akash Mahajan [1], Shivam Chaturvedi [2], Srijita Das [1], Wencong Su [2,*], Van-Hai Bui [2,*]

[1] Department of Computer and Information Science, University of Michigan-Dearborn, Dearborn, MI 48128, USA; makash@umich.edu (A.M.); sridas@umich.edu (S.D.)

[2] Department of Electrical and Computer Engineering, University of Michigan-Dearborn, Dearborn, MI 48128, USA; shivamc@umich.edu (S.C); wencong@umich.edu (W.S.), vhbui@umich.edu (V.H.B.)

* Corresponding author

**Abstract:** The selection of optimal design for power electronic converter parameters involves balancing efficiency and thermal constraints to ensure high performance without compromising safety. This paper introduces a probabilistic-learning-based stochastic surrogate modeling framework to address this challenge and significantly reduce the time required during the design phase. The approach begins with a neural network classifier that evaluates the feasibility of parameter configurations, effectively filtering out unsafe and/or impractical inputs. Subsequently, a probabilistic prediction model estimates the converter's efficiency and temperature while quantifying prediction uncertainty, providing both performance insights and reliability metrics. Finally, a heuristic optimization-based model is employed to optimize a multi-objective function that maximizes efficiency while adhering to thermal constraints. The optimization process incorporates penalty terms to discourage solutions that violate practical thresholds, ensuring actionable and realistic recommendations. An advanced heuristic optimization method is used to find the optimal solution and is compared with several well-known search algorithms, including Genetic Algorithm (GA), Particle Swarm Optimization (PSO), Simulated Annealing (SA), Tabu Search (TS), and Stochastic Hill Climbing (SHC). The results demonstrate significant improvements in predictive accuracy and optimization outcomes, offering a robust solution for advancing power electronics design.

# 1. Introduction

One of the important components in modern electronic devices is a power electronic converter (PEC), a device designed to convert power from one form to another. Power electronic converters come in various forms for different purposes, such as converting alternating current (AC) or direct current (DC) to AC or DC. Therefore, the design of PECs is a fundamental aspect of power electronics and plays a crucial role in applications ranging from household appliances to industry-grade power systems and smart grids. Regardless of their conversion type, PECs must operate with high efficiency ($\eta$) while maintaining their internal temperature ($\theta$) in harmony with the ambient temperature to ensure reliable performance and longevity. Achieving an optimal balance between efficiency and device temperature presents a significant challenge due to the complex and non-linear relationship between design parameters, efficiency, and temperature. For example, resistance in circuits leads to power loss, reducing efficiency—an undesirable outcome in high-efficiency PECs. Depending on the circuit design, there could be more than one parameter that directly or indirectly affects the performance metrics mentioned above. Even when all possible performance-affecting parameters are identified, the large and granular design parameter space makes selecting the optimal design virtually impossible and highly expensive. Simulation models can be built to explore this design space, assess system sensitivity, and perform what-if analyses. However, such experiments require thousands or more simulation instances and significant computational power to yield accurate results. This issue can be effectively addressed with the help of surrogate models [1], which approximate the behavior of such complex systems using fewer computational resources. Surrogate models can be viewed as a specialized form of supervised machine learning, where the underlying system is estimated using a data-driven, bottom-up approach that only requires sufficient data on the system's inputs and corresponding results of interest [2].

There is a plethora of studies available that has used surrogate models to solve a variety of problems in power electronics and electrical engineering. [3] provide a comprehensive overview of surrogate modeling techniques in power electronics, highlighting the use of machine learning methods such as Kriging and neural networks to replace computationally intensive simulations. In the realm of thermal management, [4] develop an artificial neural network-based surrogate model to predict the thermal behavior of power electronic modules, significantly reducing simulation time while maintaining accuracy. For signal integrity and microwave circuit applications, [5] conduct a comparative study of various surrogate modeling methods, including support vector machines and Gaussian process regression, assisting engineers in selecting appropriate modeling techniques based on specific application requirements. [6] focus on the electromechanical domain, employing artificial neural networks to model the torque of electrical machines, achieving high accuracy and demonstrating the potential of machine learning in electromechanical system design. Addressing reliability concerns, [7] introduce a surrogate model-based method for predicting the reliability of PECs by considering component degradation and failure mechanisms, enhancing predictive maintenance strategies for power systems. [8] propose an AI-driven methodology for the automated design of reliable power electronic systems, integrating surrogate models to evaluate reliability metrics and streamline the design process. Incorporating uncertainty quantification, [9] utilize support vector machines to build surrogate models that account for variability in system parameters, enhancing robustness under uncertain conditions. [10] review probabilistic surrogate modeling using Gaussian processes, emphasizing their capability to provide uncertainty estimates alongside predictions, vital for risk assessment in PEC design. [11] discuss recent advancements in surrogate modeling methods for uncertainty quantification and propagation, highlighting techniques that improve the accuracy and efficiency of uncertainty analyses in complex systems. In optimization contexts, [12] apply surrogate model-based multi-objective optimization to the design of high-speed permanent magnet synchronous machines, balancing multiple performance criteria and demonstrating the effectiveness of surrogate models in complex optimization problems. [13] present a hybrid optimization technique combining surrogate models with evolutionary algorithms for power amplifier design, achieving efficient exploration of the design space and leading to high-performance solutions with reduced computational effort. Additionally, [14] propose a general surrogate-based methodology for fast and reliable analysis and design optimization of power delivery networks, formulating a generic surrogate model methodology exploiting passive lumped models optimized by parameter extraction to fit PDN impedance profiles. Despite their advantages, surrogate models have limitations, including dependence on the quality of training data and potential inaccuracies outside the training domain. Nevertheless, their integration into PEC design processes offers significant benefits in terms of efficiency, adaptability, and the ability to handle complex, multi-parameter systems.

To address these challenges, this paper proposes a probabilistic-learning-based stochastic surrogate modeling framework that integrates classification, probabilistic regression, and heuristic optimization-based modeling to enhance and streamline the power converter design process. The framework consists of three key components: first, a neural network-based classifier evaluates the feasibility of design parameters, filtering out impractical configurations to reduce computational overhead. Second, a probabilistic regression model approximate efficiency and temperature while quantifying uncertainty, enabling more informed decision-making. Lastly, an advanced heuristic optimization-based algorithm iteratively refines design parameters by maximizing efficiency while adhering to thermal constraints. The proposed optimization process incorporates penalty terms to prevent impractical solutions while comparing heuristic-based techniques with conventional physics-based methods. By incorporating uncertainty estimation and feasibility constraints, this framework provides a robust, data-driven solution for advancing power converter design and optimization. To address these challenges, this paper proposes a probabilistic learning-based surrogate modeling framework for optimizing power electronic converter (PEC) parameters. The framework combines a neural network-based feasibility classifier, a probabilistic regression model for jointly predicting efficiency and temperature, and a suite of stochastic optimization algorithms that minimize a fitness function under practical constraints. Unlike conventional methods, this approach integrates uncertainty quantification and feasibility checks directly into the optimization process, enabling robust and efficient exploration of the design space.

The key contributions of this work include:

1) Development of a high-accuracy neural network classifier to filter out infeasible designs early in the pipeline.
2) Implementation of a probabilistic surrogate model using Natural Gradient Boosting for accurate and uncertainty-aware predictions of converter performance metrics.
3) An advanced genetic algorithm is designed using a multi-objective fitness function with a soft penalty-reward mechanism that incorporates classification confidence and prediction intervals to achieve more robust optimization; and
4) An extensive comparative analysis of heuristic optimization algorithms in terms of convergence speed, runtime efficiency, and final design quality.

The rest of the paper is organized as follows. Section 2 describes the converter simulation setup and outlines the data generation process used to create a comprehensive design dataset. Section 3 introduces the proposed surrogate modeling framework, detailing the classification, regression, and optimization components. Section 4 presents experimental results and performance evaluations of the different modeling and optimization techniques. Section 5 concludes the paper by summarizing the main findings, and Section 6 discusses potential directions for future research.

## 2. Converter Simulation and Data Generation

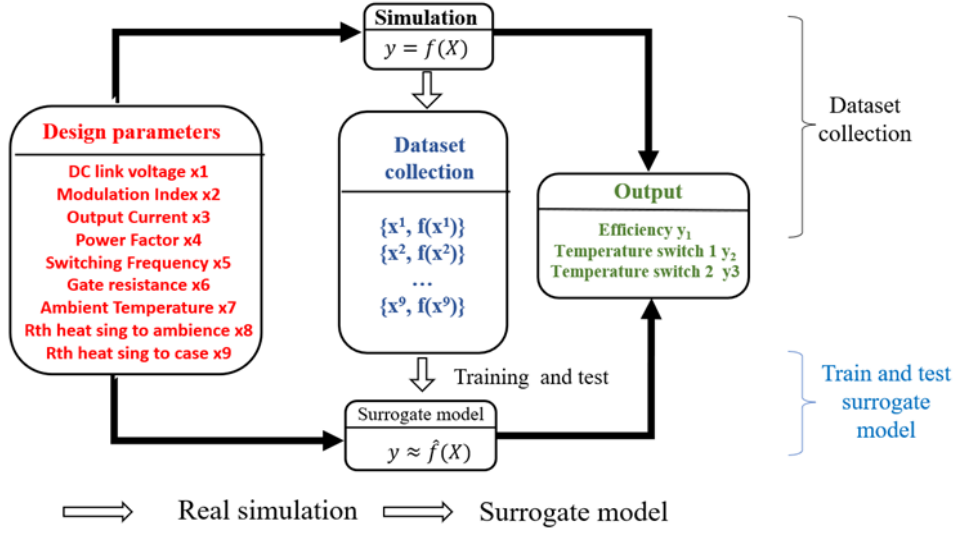

**Figure 1.** Simulation and Data Generation

A half bridge converter is used to validate the optimal parameters selection for design process. A typical controlled half-bridge converter consists of two active switches which are switched at a particular switching frequency, two capacitors which divide the input voltage into two equal half, and a load. Different forms of half bridge converter have found a wide range of applications like motor applications, inverters, switched mode power supplies, household equipment, renewable energy systems, and industrial modules. The components of a half-bridge converter need to be chosen to achieve the optimal performance. The switch of the half-bridge converter consists of an IGBT or MOSFET, which consists of series resistance, which affects the loss and temperature rise with respect to the switching frequency and load current which passes through it during a typical operating point. Furthermore, different operating conditions such as modulation index and load power factors significantly affect the losses and rise in temperature of the switches. The higher modulation index delivers more power to the output terminals, while increasing the losses through the switches. A load with a poor power factor leads to higher current through the switches, which also leads to an increased temperature and increased losses. This can be observed from a typical relation which is as follows:

$$I_o = \frac{mV_{dc}}{2R_L}, \qquad I_{or} = \frac{mV_{dc}}{2\sqrt{2}R_L} \qquad (1)$$

In terms of power converter,

$$I_{orf} = \frac{mV_{dc}k_{pf}}{2\sqrt{2}R_L}, \qquad I_{in} = \frac{P}{V_{in}k_{pf}} \tag{2}$$

where, $I_{in}$ is the input current from the source side of the half bridge converter, $V_{dc}$ is the source DC voltage, $k_{pf}$ is the power factor, $R_L$ is the load connected at output terminals, $I_o$ is the current through the output terminals, $I_{or}$ is the RMS output voltage, $I_{orf}$ is the output current with respect to the power factor, $P$ is the active power delivered to the load, and $m$ is the modulation index. These equations show a critical relation between the input output and operating parameters. Hence, these conditions must be designed properly, and the switches must be selected to achieve the desired efficiency and must operate within the desired temperature level. This is needed because the switches are designed to operate within a temperature range, and higher temperatures might lead to higher losses and failure. The characteristics of IGBT or MOSFETs are highly dependent on the temperature. Furthermore, gate resistance plays a critical role in switching characteristics of the switches. A higher value of gate resistance leads to slower switching which further leads to lower current and voltage changes with respect to time, but it leads to higher losses. On the other hand, a low value of gate resistance leads to faster switching but higher electromagnetic interference and ringing. In terms of heat dissipation of the half bridge converter, thermal resistance is an important quantity. A higher thermal resistance leads to lower heat dissipation which leads to higher rise in temperature, on the other hand, a lower resistance leads to higher heat dissipation, and reduction in the temperature of the converter. Higher thermal resistance causes a significant increase in junction temperature. A proper heat sink orientation and placement ensures efficient heat dissipation, preventing hot spots on the device.

The data for which is used for present work is generated by simulating a half bridge converter in various operating scenarios and observing the effect on temperature rise and efficiency for each operating condition. Different parameters of interest consist of: dc link voltage ($x_1$), amplitude of the modulation index($x_2$), output current amplitude ($x_3$), output power factor ($x_4$), switching frequency ($x_5$), gate resistance ($x_6$), ambient temperature ($x_7$), thermal resistance heat sink to ambiance ($x_8$), thermal resistance heat sink to case ($x_9$). All these parameters are varied with a pre-defined range and corresponding outputs are observed. The simulation is run at a fixed time step of ten microseconds and the readings are observed after steady state condition is reached. A typical process of data generation is presented in Figure 1.

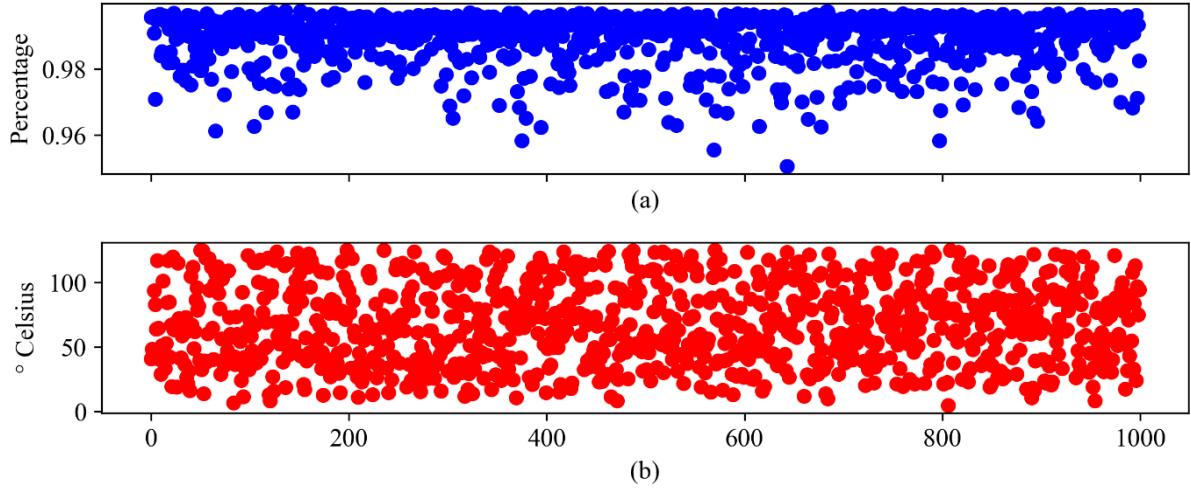

**Figure 2.** Visualization of thousand, randomly selected observations of (a) efficiency and (b) temperature

## 3. Proposed System Design

This section begins with a brief introduction of the data that was used to design the surrogate system. Following the data, the neural network that serves as a design classifier is described. We have used various regression algorithms to probabilistically predict efficiency and temperature resulting from the given configuration of the design parameters. Moving further, design optimization algorithms were described, and evaluation metrics are formularized with their purpose.

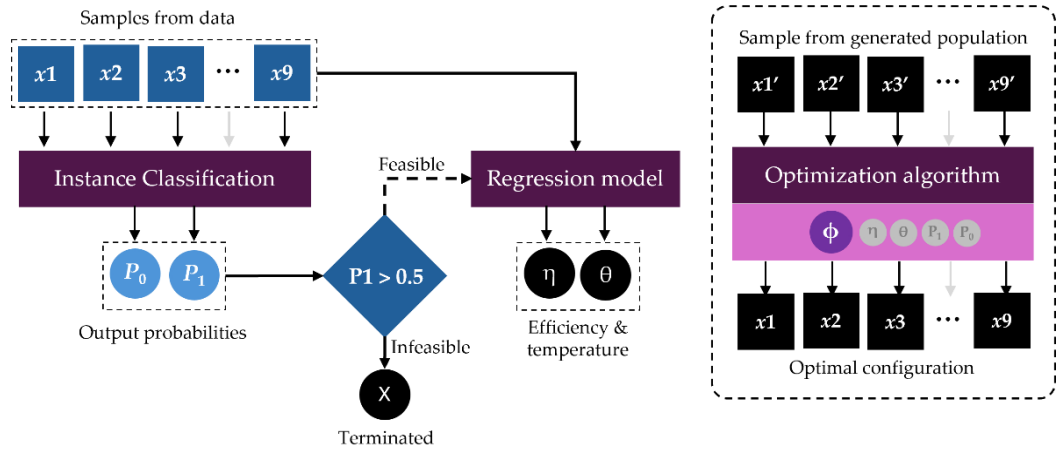

**Figure 3.** Block diagram of machine learning pipeline for proposed surrogate model

### 3.1. Design parameters dataset for classification and probabilistic prediction

The vital component of the proposed surrogate system is the design dataset. The selected design dataset consists of nine input parameters, namely $x_1, x_2, x_3, \ldots, x_9$ and two resulting quantities that indicate the status of the power converter, i.e. efficiency and temperature, which are represented in the dataset as $y_1$ and $y_2$, respectively. There are a total of thirty thousand samples in the selected dataset. However, the dataset is slightly biased.

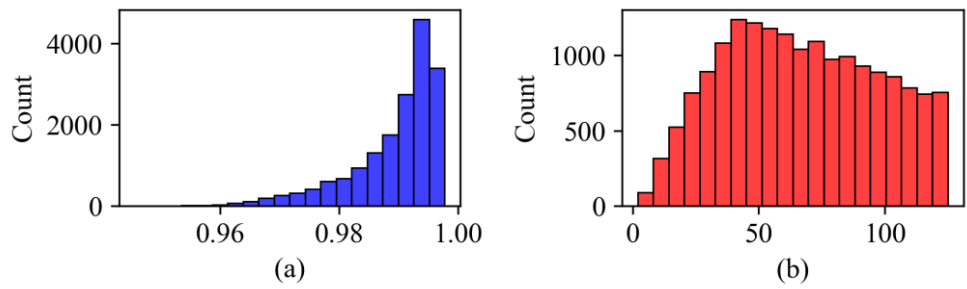

**Figure 4**. Frequency distribution of the (a) efficiency and (b) temperature

Upon visualization and analysis of the design parameters, it was observed that the efficiency $y_1$ has a very narrow distribution to the right. The basic statistics of the dataset are shown in Table X. This bias can potentially jeopardize the performance of all the system components like classification, regression, and optimization. To address the issue of narrow distribution, we have applied a simple transformation in the form of Z standardization across all features $x_1, x_2, x_3, \ldots, x_9$ and targets $y_1$ and $y_2$ to ensure that all values across each feature and target have a similar scale and avoid the computation complexity for machine learning algorithms. This simple transformation has significantly improved the performance for regression and classification models compared to other transformation techniques evaluated (see Appendix A.1). This dataset has multiple purposes in our proposed surrogate model, such as (a) developing a classification model that tests the feasibility of the design based on the resulting quantities such as efficiency and temperature and (b) developing a regression model to probabilistically estimate these quantities. When developing an optimization model, the objective function ensures that the selected design is within the range of feature values, is feasible, and results in optimal quantities given operational environment constraints. Therefore, the proposed system works in amalgamation of classification, regression, and optimization with the major component being the input dataset. From now on, efficiency and temperature together will be referred to as 'quantities'. As briefly discussed previously, our proposed surrogate system has multiple components that work synchronously to produce optimal parameters that will aid the design process of a power converter. The following subsections will provide a detailed description of each component consisting of its general importance in the surrogate model, input and output specification, and associated algorithms. The block diagram of machine learning pipeline is visualized in Figure 2.

## 3.2. Classification of design parameters for feasibility

The objective of the instance classification component is to assign each design, defined by the fore-mentioned nine parameters, to one of two classes: feasible or not feasible. This is critical because the data set contains designs that are impractical for real-world operations; for example, designs may produce temperatures exceeding 125° C, which is beyond the physical capacity of the power converter. Consequently, classification models have been trained to identify such designs solely on the basis of their parameters. Additionally, each design is manually labeled according to the constraint that efficiency must range from 0 to 1 and temperature must not exceed 125° C. Designs that do not meet these criteria are flagged accordingly. The classification probabilities are provided to offer the design optimization algorithm an indication of the confidence associated with each classification. In cases where the model produces an incorrect result (e.g., a false positive or a false negative), a confidence threshold is used to apply an appropriate penalty factor. As a result, even if a design is classified as infeasible, it may still be considered for quantity estimation based on the confidence level of the dominant class, ensuring a thorough exploration of the design parameter space.

## 3.3. Probabilistic prediction of efficiency and temperature

Based on the nine design parameters, a probabilistic regression algorithm is employed to estimate key quantities, specifically efficiency and temperature. For accurate predictions, it is imperative that the probabilistic regression model achieves high predictive accuracy while maintaining appropriate confidence levels. To address this, we evaluated various probabilistic multivariate regression models that are capable of estimating the correlation between these variables and predicting them concurrently. It is important to note that these probabilistic models were not trained on the entire dataset, which includes both feasible and infeasible designs. This decision was made to ensure that the quantity estimations are as dependable as possible, thereby excluding all infeasible designs from the training process. Furthermore, all optimization objectives are solely based on the upscaled values of efficiency and temperature. Objectives that incorporate the prediction interval utilize the mean and variance of the predicted distribution to calculate a penalty factor. Consequently, the regression model must be capable of mapping each input design to various potential outcomes for efficiency and temperature, thereby facilitating the determination of a precise fitness value for each unique design.

The following are the probabilistic regression algorithms compared for the quantity estimation:

**Gaussian Process Regression [15]**. A non-parametric Bayesian approach used for regression tasks that provides not only predictions but also an estimation of uncertainty. Unlike traditional regression models that assume a fixed functional form, GPR defines a distribution over functions, enabling flexible modeling of complex, nonlinear relationships. A Gaussian process is fully characterized by a mean function $m(x)$ and a covariance function (kernel) $k(x, x')$ where the kernel function captures the similarity between data points and dictates the smoothness of predictions.

**Natural Gradient Boosting [16].** An advanced probabilistic machine learning technique that extends traditional gradient boosting by incorporating uncertainty estimation through the use of natural gradients in probabilistic space. Unlike standard gradient boosting methods such as XGBoost or LightGBM, which minimize a deterministic loss function, NGBoost optimizes full probability distribution by modeling both the predicted mean and variance. This is achieved by leveraging the natural gradient descent, which improves optimization efficiency in probabilistic models by operating in the Riemannian space of distributions rather than the standard Euclidean space. NGBoost models a parametric likelihood function (e.g., Gaussian, Laplace, or Cauchy) and iteratively refine its predictions by adjusting both the central tendency and dispersion of the target distribution.

**Monte Carlo Neural Network [17].** This method leverages Monte Carlo (MC) sampling techniques to introduce probabilistic inference and uncertainty quantification into deep learning models. Unlike deterministic neural networks, which provide point estimates, MCNNs estimate the distribution of outputs by performing multiple stochastic forward passes. This is commonly achieved through Monte Carlo Dropout (MC Dropout), where dropout layers remain active during inference, effectively sampling different subnetworks and yielding a distribution over predictions. This approach approximates Bayesian inference in deep neural networks without explicitly maintaining a full posterior distribution over weights.

**Bayesian Neural Network (BNN) [18]**. This approach extends traditional feed-forward neural networks by placing probability distributions over the network's weights and biases, enabling principled uncertainty quantification in predictions. During training, weight and bias parameters are not treated as fixed values but are randomly sampled from prior distributions, typically Gaussian. The true posterior over these parameters is intractable to compute directly due to high dimensionality, so variational inference is employed: a tractable approximate ("variational") posterior—also Gaussian—is optimized to minimize the Kullback–Leibler (KL) divergence from the true posterior. Once trained, predictions are made by sampling weights multiple times and averaging the resulting forward passes, which produces a predictive distribution rather than a single point estimate. This allows the model to capture both aleatoric variability and epistemic uncertainty in energy demand forecasts

**3.4. Optimizing the design parameters**

The previous two components work synchronously to obtain the desired quantities based on the user-provided input. However, in addition to these component we introduce an additional component that can help estimate the optimal design given a desired set of quantities. We have evaluated various stochastic optimization methods to iteratively optimize a given randomly initialized design based on the desired output quantities given by the user. Instead of just focusing on optimizing a single fitness function, we have tested following classes of functions for efficiency and temperature namely (1) Harsh Penalty for infeasible solution, (2) Soft penalty for infeasible solution with soft reward for feasible solution (using confidence in classification), and (3) Prediction-interval based estimation of efficiency/temperature with soft penalty/reward function. All these functions were designed for both single objective and multi-objective. Models based on neural networks are trained and evaluated using PyTorch framework.

**3.4.1. Proposed Multi-Objective Genetic Algorithm**

Genetic Algorithm (GA) [19] is a population-based evolutionary optimization technique inspired by the principles of natural selection and genetics. It operates by evolving a population of candidate solutions over successive generations using operations such as selection, crossover, and mutation. Each individual in the population represents a potential solution, and its quality is assessed using a fitness function. The algorithm selects the fittest individuals as parents and combines their characteristics through crossover to produce offspring, while mutation introduces small random changes to maintain genetic diversity and avoid premature convergence. This population-level exploration allows GA to effectively navigate large and complex search spaces, making it well-suited for multi-dimensional optimization problems. In this study, GA is applied to optimize nine design parameters with the aim of achieving high efficiency and moderate temperature levels, ideally close to the ambient reference. For this purpose, the objective function is a very essential component which ensures a balanced multi-objective optimization. It turns out that objective function should also be able to distinguish between feasible solutions and infeasible solutions. Therefore, when defining the objective function, we incorporate a design classifier and a probabilistic prediction model.

**The objective function:** The objective function for GA is a multi-objective function with a soft penalty mechanism based on classification confidence and probabilistic predictions. The function first evaluates the feasibility of a candidate solution using a classification model, returning a confidence score. Then, a probabilistic regression model estimates the 95% prediction interval for the target quantity (efficiency or temperature). The objective function penalizes infeasible solutions and rewards feasible ones proportionally to classification confidence, ensuring a smooth optimization landscape rather than hard constraint rejection. This structure improves optimization convergence by allowing controlled exploration of infeasible regions. The function maintains high quality by integrating uncertainty-aware predictions, reducing the risk of overfitting to misleading deterministic estimates. Despite incorporating multiple components, it remains computationally efficient and modular, making it adaptable across various optimization techniques while maintaining interpretability and robustness. Algorithm 1 shows the steps in evaluating a solution and obtaining fitness value.

| Algorithm 1: A multi-objective fitness function with soft penalty mechanism | | |
|---|---|---|
| **Input:** | Candidate solution, X; goal temperature, t; | |
| **Data:** | Step 1: | Pass X to classifier and obtain classification confidence, $P$. |
| | Step 2: | Pass X to probabilistic regression model to obtain prediction mean, $\mu$, and prediction standard deviation, $\sigma$. |
| **Evaluation:** | Step 1: | Declare penalty factor $PF = 5$ |
| | Step 2: | Obtain upper and lower bound for efficiency, **Y1**, and temperature, **Y2** using the $\mu$ and $\sigma$ |
| | Step 3: | Assign **Y1** and **Y2** a randomly selected value between corresponding upper and lower bound |
| | Step 4: | Multiply efficiency by 100 to scale up from 0-1 range |
| | Step 5: | Calculate $F = |100 - Y1|^2 + |t - Y2|^2$ |
| | Step 6: | Apply penalty based on the classification confidence $F = F + PF \times P$ |
| **Output:** | Fitness value, **F** | |

The same objective function is used with other optimization algorithms for a fair comparison. The following is the description of the implementation of the GA algorithm:

**Chromosome representation:** Each candidate solution is represented as a real-valued vector $X = [x_1, x_2, x_3, \ldots, x_n]$ where $n$ denotes the number of design parameters. This is encapsulated within the Chromosome class, which also enforces lower and upper bounds for each gene. The enforcement mechanism ensures that any gene modification due to mutation or crossover remains within its predefined design limits, preserving the validity of solutions throughout the evolutionary process.

**Mutation operation:** Mutation introduces random variations into the genetic structure of each chromosome. For every gene $x_i$, a random number is sampled, and if it falls below a mutation threshold $p_m$, the gene is perturbed by a fixed scalar $\alpha$. This perturbation is then passed through a bounding function to ensure it remains within the specified domain. This operation allows the algorithm to maintain genetic diversity and explore new areas of the design space.

**Crossover operation:** Crossover is implemented using a single-point recombination strategy. A random crossover point is selected along the gene vector, and two offspring are generated by exchanging segments of two parent chromosomes at that point. The resulting children inherit a mix of features from both parents, promoting information sharing across the population. The gene-wise bounds are also recombined to ensure that the offspring remain feasible under their respective parameter constraints.

**Population initialization:** The population is initialized by sampling each gene of every chromosome uniformly within its allowable range. This operation is executed in the constructor of the Population class and generates a diverse set of candidate solutions. The diversity in the initial population is essential for avoiding premature convergence and for enabling a broad exploration of the design space during early generations.

**Selection mechanism:** Selection is performed using inverse fitness-proportional sampling, commonly known as roulette wheel selection. Each chromosome's probability of being selected is inversely proportional to its fitness score, favoring better-performing individuals while maintaining stochasticity. An epsilon term is added to avoid division by zero and ensure numerical stability during probability computation.

**Elitism and replacement:** To prevent the loss of high-quality solutions during evolutionary operations, elitism is applied. The best-performing chromosome from the current generation is preserved and injected into the next generation, replacing a randomly chosen individual. This ensures that the overall best solution found so far is always retained, promoting convergence stability across generations.

**Genetic algorithm execution:** The optimization process is governed by the ga() function, which performs population initialization, iterative fitness evaluation, and evolution over a fixed number of generations. In each generation, chromosomes are evaluated, selected, recombined, mutated, and updated. The best-performing chromosome from each generation is tracked, and both the fitness progression and final solution set are returned upon completion for further analysis or deployment.

Alongside GA, the following are the optimization algorithms compared for design optimization:

**Stochastic Hill Climbing (SHC) [20].** SHC is a local search optimization method that iteratively refines a solution by applying small random perturbations and accepting changes based on their effect on the objective function. Unlike deterministic hill climbing, SHC introduces randomness in candidate selection, allowing broader exploration and reducing the risk of getting trapped in local optima. It may also accept worse solutions with a certain probability to escape plateaus or deceptive regions. In this study, a multi-objective SHC algorithm is employed to optimize nine design parameters for high efficiency and moderate temperature. The optimization is guided by a custom objective function that integrates a design classifier and a probabilistic prediction model, enabling soft penalization of infeasible solutions and uncertainty-aware evaluation.

**Particle Swarm Optimization [21].** It is a population-based method of optimization algorithm inspired by the collective behavior of birds and fish schools. It explores the search space using a swarm of particles, where each particle represents a candidate solution. Each particle updates its position based on its own best-known position (personal best, $P_{best}$) and the globally best-known position found by any particle (global best, $g_{best}$). The update follows the velocity equation:

$$v_i(t+1) = wv_i(t) + c_1 r_1 (P_{best} - x_i) + c_2 r_2 (g_{best} - x_i) \qquad \ldots(3)$$

where $w$ is the inertia weight, $c_1$ and $c_2$ are acceleration coefficients, and $r_1$, $r_2$ are random numbers.

**Simulated Annealing [22].** Simulated Annealing (SA) is a probabilistic optimization algorithm inspired by the annealing process in metallurgy, where a material is gradually cooled to reach a low-energy crystalline state. SA is particularly effective for combinatorial and non-convex optimization problems where traditional methods struggle with local minimum. The algorithm starts with an initial solution and explores the search space by iteratively making small random changes. A new solution is accepted if it improves the objective function; otherwise, it is accepted with a probability defined by the Boltzmann distribution:

$$P = e^{-\frac{\Delta E}{T}} \tag{4}$$

where $\Delta E$ is the change in objective function value and $T$ is the current temperature. As the temperature gradually decreases, SA transitions from exploratory (accepting worse solutions) to exploitative (refining the best-found solutions). This cooling schedule prevents premature convergence.

**Tabu Search [23].** It is a metaheuristic optimization algorithm designed to escape local optima by maintaining a short-term memory structure known as the tabu list. Unlike traditional local search algorithms, which can become trapped in suboptimal solutions, TS allows moves to worse solutions while avoiding cycles by prohibiting recently visited states. The search process iteratively explores a neighborhood of candidate solutions and selects the best non-tabu solution, unless a tabu move leads to a globally improved solution, in which case the tabu restriction is overridden. The tabu list dynamically updates, storing restricted moves for a fixed number of iterations before allowing them again. Additionally, aspiration criteria are used to permit high-quality moves even if they are tabu.

### 3.5. Evaluation Metrics and experimental settings

Depending on the component, various evaluation metrics were used to select the best-performing algorithm. Table 1 outlines component-wise evaluation metrics used in this study.

**Table 1.** Component-wise evaluation metrics

| Evaluation metric | Formula | Eq. No | Purpose |
|---|---|---|---|
| **Classification1** | | | |
| Binary Cross-Entropy (BCE) | $-\frac{1}{N}\sum_{i=1}^{N} y_i \times +log(1 - y_i) \times log(1 - P(y_i))$ | (6) | Measures the error between predicted probabilities and true labels, useful for evaluating the confidence of binary classification models. |
| Precision | $\frac{TP}{TP + FP}$ | (7) | Evaluates the ratio of true positives to the sum of true and false positives, useful in contexts where the cost of a false positive is high. |
| F1 score (F1) | $\frac{TP}{TP + \frac{1}{2}(FP + FN)}$ | (8) | Balances precision and recall, providing a single metric for model accuracy in cases with |
| | | | uneven class distribution, especially when false negatives and positives are costly. |
| Accuracy | $\frac{TP + TN}{TP + FN + TN + FP}$ | (9) | Measures the proportion of true results (both true positives and true negatives) among the total number of cases examined, appropriate for balanced datasets and general classification efficacy. |
| **Probabilistic Prediction** | | | |
| Root Mean Squared Error (RMSE) | $\sqrt{\frac{\sum_{i=1}^{N}(y_i - \hat{y}_i)^2}{N}}$ | (10) | Quantifies the square root of the average squared differences between predicted and actual outcomes, highlighting model performance and sensitivity to large errors in regression models. |
| Mean Absolute Error (MAE) | $\frac{\sum_{i=1}^{N}(y_i - \hat{y}_i)}{N}$ | (12) | Measures the average magnitude of errors in a set of predictions, without considering their direction, useful for understanding average outcome deviations in regression models. |
| Mean Absolute Percentage Error (MAPE) | $\frac{1}{N}\sum_{i=1}^{N}\left|\frac{y_i - \hat{y}_i}{y_i}\right| \times 100$ | (13) | Expresses accuracy as a percentage, and it measures the average absolute percent error per forecasted value; it's particularly useful when dealing with scale-independent error measurements. |
| Prediction Interval Coverage Probability (PICP) | $\frac{1}{N}\sum_{i=1}^{N} c_i, \quad c_i = \begin{cases} 1, & L_i \leq y_i \leq U_i \\ 0, & otherwise \end{cases}$ | (14) | Measures the proportion of times that actual outcomes fall within a forecasted interval, indicating the reliability and calibration of predictive uncertainty estimates. |
| Continuous Ranking Probability Score (CRPS) | $\int_{-\infty}^{\infty} \left(F(y) - 1(y \geq y_{obs})\right)^2 dy$ | (15) | Assesses the accuracy of probabilistic forecasts; it is particularly useful when |

| | | | evaluating the quality of the entire probability distribution of forecasted values. |
|---|---|---|---|
| R2 Score (R2) | $1 - \frac{\sum_{i=1}^{N}(y_i - \hat{y}_i)^2}{\sum_{i=1}^{N}(y_i - \bar{y}_i)^2}$ | (16) | Measures the proportion of variance in the dependent variable that is predictable from the independent variables, making it useful for assessing the explanatory power of regression models. |
| Negative Log-Likelihood (NLL) | $-\sum_{i=1}^{N} \log p(y_i\|\theta)$ | (17) | Provides a measure of how well a model predicts the actual class labels with probabilistic outputs, focusing on the probability assigned to the true classes across the dataset. |

**Experimental settings.** All experiments are run on standard computing environment with Python 3.9 and CUDA support. The ANN classifier is built using PyTorch and is trained for 150 epochs with a learning rate of 0.001 and Adam optimizer. Similarly, MCNN and BNN models are built using PyTorch and trained on 100 epochs with a learning rate of 0.001 and Adam optimizer. All the classic classifiers like decision trees, random forests, and logistic regression are imported from Scikit-learn library, a popular Python library for machine learning. NGB and XGBoost are imported from their corresponding special packages. All optimization algorithms are written from scratch to support the multi-objective fitness function, CUDA-enabled classifier, and probabilistic regressor. Exact configuration of all optimization algorithms is listed in Table 5.

## 4. Results

### 4.1. Comparative performance of neural network and classical classifiers

Table 2 presents the performance comparison of various classifiers trained using five-fold cross-validation. Each model was trained and validated across five folds to ensure consistent and robust evaluation. The table reports both average and best-case performance metrics, including BCE and accuracy.

Table 2. Performance comparison of classification algorithms using five-fold cross-validation

| Fold | Decision Tree | | XGBoost | | Random Forest | | Logistic Regression | | ANN | |
|---|---|---|---|---|---|---|---|---|---|---|
| | BCE | Acc. | BCE | Acc. | BCE | Acc. | BCE | Acc. | BCE | Acc. |
| 1 | 2.469 | 0.931 | 0.054 | 0.975 | 0.126 | 0.962 | 0.097 | 0.957 | 0.033 | 0.988 |
| 2 | 2.485 | 0.931 | 0.050 | 0.975 | 0.118 | 0.966 | 0.099 | 0.954 | 0.034 | 0.991 |
| 3 | 2.469 | 0.931 | 0.049 | 0.977 | 0.122 | 0.965 | 0.098 | 0.952 | 0.033 | 0.993 |
| 4 | 2.639 | 0.926 | 0.050 | 0.979 | 0.125 | 0.960 | 0.112 | 0.949 | 0.038 | 0.991 |
| 5 | 2.802 | 0.922 | 0.055 | 0.974 | 0.125 | 0.962 | 0.102 | 0.952 | 0.036 | 0.991 |
| Avg. | 2.573 | 0.928 | 0.052 | 0.976 | 0.123 | 0.963 | 0.101 | 0.952 | 0.035 | 0.991 |
| Best | 2.469 | 0.931 | 0.049 | 0.979 | 0.118 | 0.966 | 0.097 | 0.957 | 0.033 | 0.993 |

Among all classifiers, the ANN achieved the best performance in terms of both BCE and Accuracy. XGBoost also demonstrated competitive results, with only marginal differences in average metrics. The confusion matrices (Figure X) indicate that ANN maintained lower false positive rates compared to XGBoost across most folds.

Table 3. Test performance comparison of various classification models

| Model | Accuracy | F1 score | Precision | Recall | AUC-PR | BCE |
|---|---|---|---|---|---|---|
| **Decision Tree** | 0.931 | 0.936 | 0.934 | 0.938 | 0.910 | 2.466 |
| **XGBoost** | 0.984 | 0.985 | 0.983 | 0.986 | 0.999 | 0.040 |
| **Random Forest** | 0.967 | 0.970 | 0.967 | 0.972 | 0.997 | 0.113 |
| **Logistic Regression** | 0.956 | 0.959 | 0.956 | 0.962 | 0.995 | 0.096 |
| **ANN** | 0.995 | 0.995 | 0.994 | 0.997 | 0.999 | 0.029 |

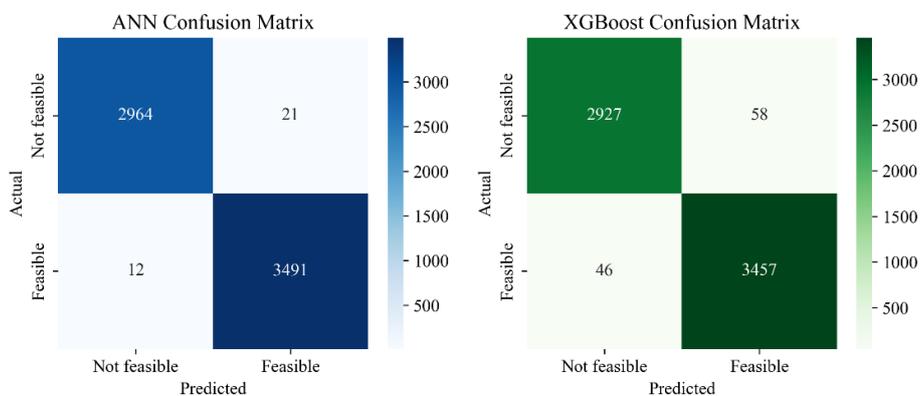

**Figure 5.** Confusion matrix for ANN (left) and XGBoost (right) classifiers

Figures 6(a) and 6(b) illustrate the learning curves for ANN, showing consistent improvements in training and validation MSE loss and accuracy across all folds, respectively. These trends suggest stable convergence behavior during training.

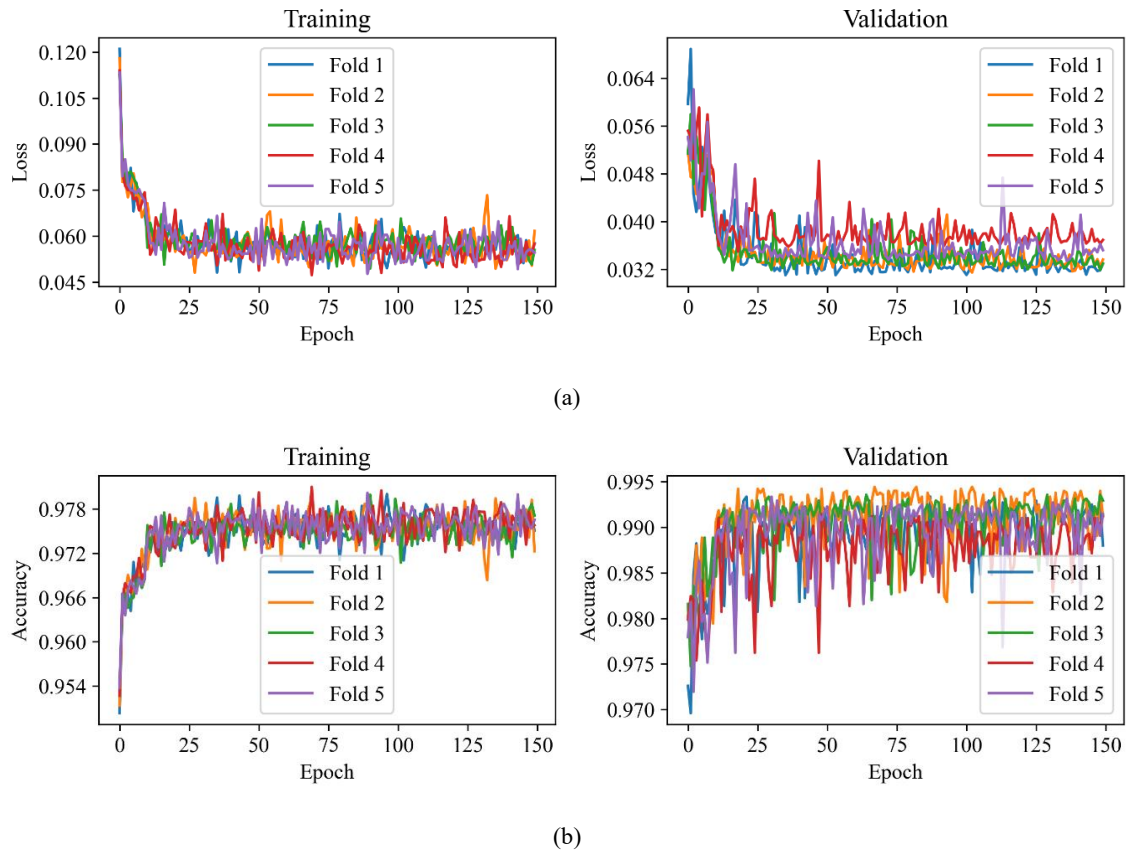

**Figure 6.** ANN learning curve for (a) MSE loss and (b) accuracy

## 4.2. Comparison of probabilistic regression algorithms for predicting efficiency and temperature

Table 4 shows the comparison of various probabilistic algorithms based on various pointwise and probabilistic metrics. As previously mentioned, all algorithms model the efficiency and temperature jointly. All algorithms have similar performance for modeling efficiency. However, NGB algorithm excels in modeling both of them.

**Table 4.** Evaluation of probabilistic models using pointwise accuracy and uncertainty quantification metrics

| Algorithm | Model | Test metrics | | | | |
|---|---|---|---|---|---|---|
| | | **RMSE** | **R2** | **MPIW** | **PICP** | **CRPS** |
| NGB (ET) | Efficiency | 0.02 | 1.00 | 0.04 | 0.95 | 0.01 |
| | Temperature | 4.95 | 0.97 | 15.95 | 0.96 | 2.58 |
| GPR (RBF) | Efficiency | 0.07 | 0.99 | 0.14 | 0.80 | 0.03 |
| | Temperature | 1.26 | 1.00 | 6.99 | 0.96 | 0.66 |
| MCNN | Efficiency | 0.11 | 0.98 | 0.80 | 0.98 | 0.07 |

|     | Temperature | 4.61 | 0.98 | 31.38 | 1.00 | 2.77 |
|-----|-------------|------|------|-------|------|------|
| BNN | Efficiency  | 0.02 | 1.00 | 0.05  | 0.93 | 0.01 |
|     | Temperature | 1.39 | 1.00 | 3.09  | 0.79 | 0.71 |

| Algorithm | NGB (ET) | GPR (RBF) | MCNN | BNN |
|-----------|----------|-----------|------|------|
| **NLL**   | -0.44    | 0.31      | 1.27 | -0.56 |

During training, all models were trained and validated. The learning curve of NGB model is shown in Figure 7 outlining the consistent learning process.

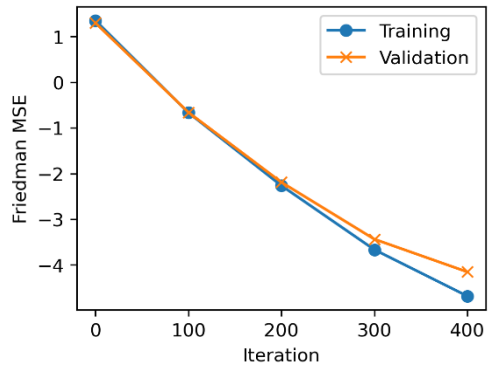

**Figure 7.** Learning curve for NGB over five hundred iterations

In the following Figure 8, hundred randomly selected points are shown from test dataset with their observed value and corresponding predicted value by NGB. Figure 8 shows prediction for efficiency and temperature, respectively. The upper and lower confidence interval (CI) is visualized for each point in the form of error bars. Shorter this CI, higher the confidence in the prediction.

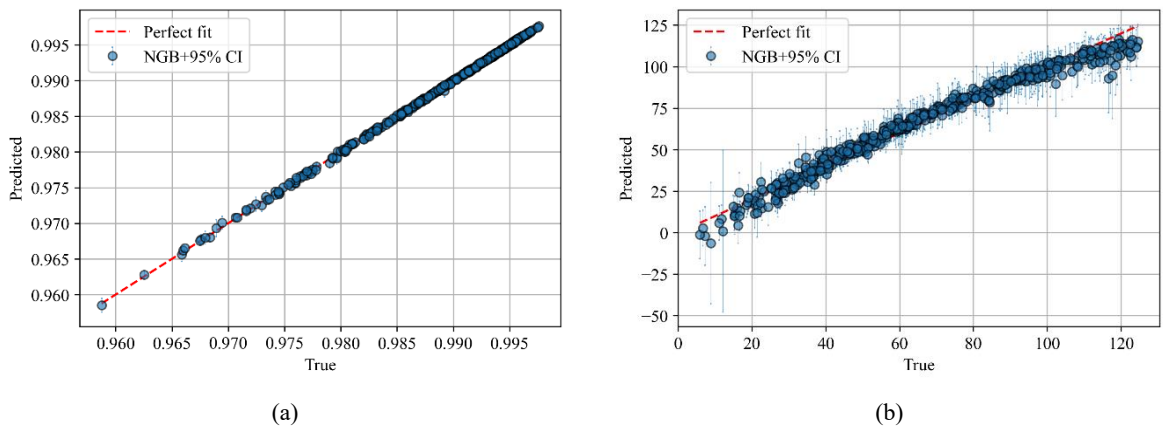

(a)          (b)

**Figure 8.** Visualization of five hundred randomly selected predictions for (a) efficiency and (b) temperature

## 4.3. Estimating optimal design parameters using stochastic optimization

Table 5 presents the hyperparameters and a summary of results for each stochastic optimization algorithm. The table highlights key performance metrics including: (1) total runtime to reach an optimal solution or complete the iterations (formatted as mm:ss), (2) the minimum fitness value achieved, and (3) the optimal efficiency and temperature values obtained.

Table 5. Hyperparameter and result summary for design optimization algorithms

| Algorithm | Parameter name | Value | Time to optimize (mm:ss) | Optimal value obtained | | |
|---|---|---|---|---|---|---|
| | | | | Fitness | Efficiency | Temperature |
| **Genetic Algorithm (GA)** | Population size | 20 | 26:01 | 0.00004 | 0.9911 | 28.00 |
| | Crossover rate | 0.4 | | | | |
| | Mutation rate | 0.3 | | | | |
| | Number of generations | 100 | | | | |
| **Particle Swarm Optimization (PSO)** | Swarm size | 20 | 6:80 | 0.00019 | 0.9904 | 28.01 |
| | Inertia weight | 0.1 | | | | |
| | Cognitive coefficient | 1.0 | | | | |
| | Social coefficient | 0.2 | | | | |
| | Number of iterations | 100 | | | | |
| **Simulated Annealing (SA)** | Temperature | 100 | 00:72 | 0.52636 | 0.9912 | 29.14 |
| | Cooling rate | 0.9 | | | | |
| | Acceptance probability | Random | | | | |
| | Number of iterations | 100 | | | | |
| **Tabu Search (TS)** | Tabu list size | 20 | 14:71 | 4.49218 | 0.9956 | 31.34 |
| | Neighborhood size | 20 | | | | |
| | Number of iterations | 100 | | | | |
| **Stochastic Hill Climbing** | Number of iterations | 100 | 01:82 | 0.02523 | 0.9916 | 27.75 |

Figure 9 illustrates the optimization progress for each algorithm across iterations. As previously discussed, the objective is to minimize the fitness function, thereby achieving optimal values for efficiency and temperature.

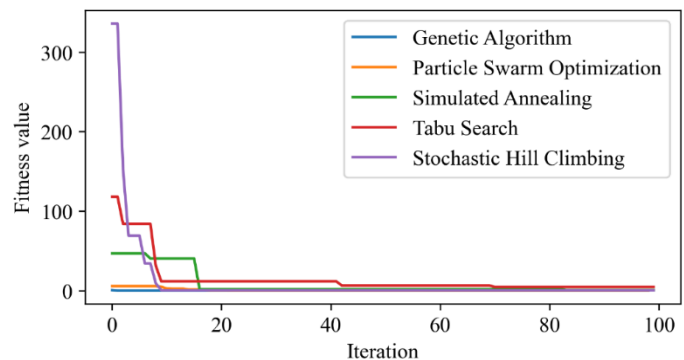

**Figure 9.** Optimization curve summary for all algorithms

The visual and tabular results together indicate variable performance across the algorithms. Among the evaluated methods, the GA algorithm demonstrates superior performance in terms of both computational efficiency and convergence speed. Specifically, Figure 10(a) shows the evolution of efficiency (left) and temperature (right) throughout the optimization process. This visualization provides insight into the dynamic behavior of the algorithm as it converges to a stable solution. Similar patterns of convergence are observed for each design parameter across iterations, reinforcing the effectiveness of the optimization process.

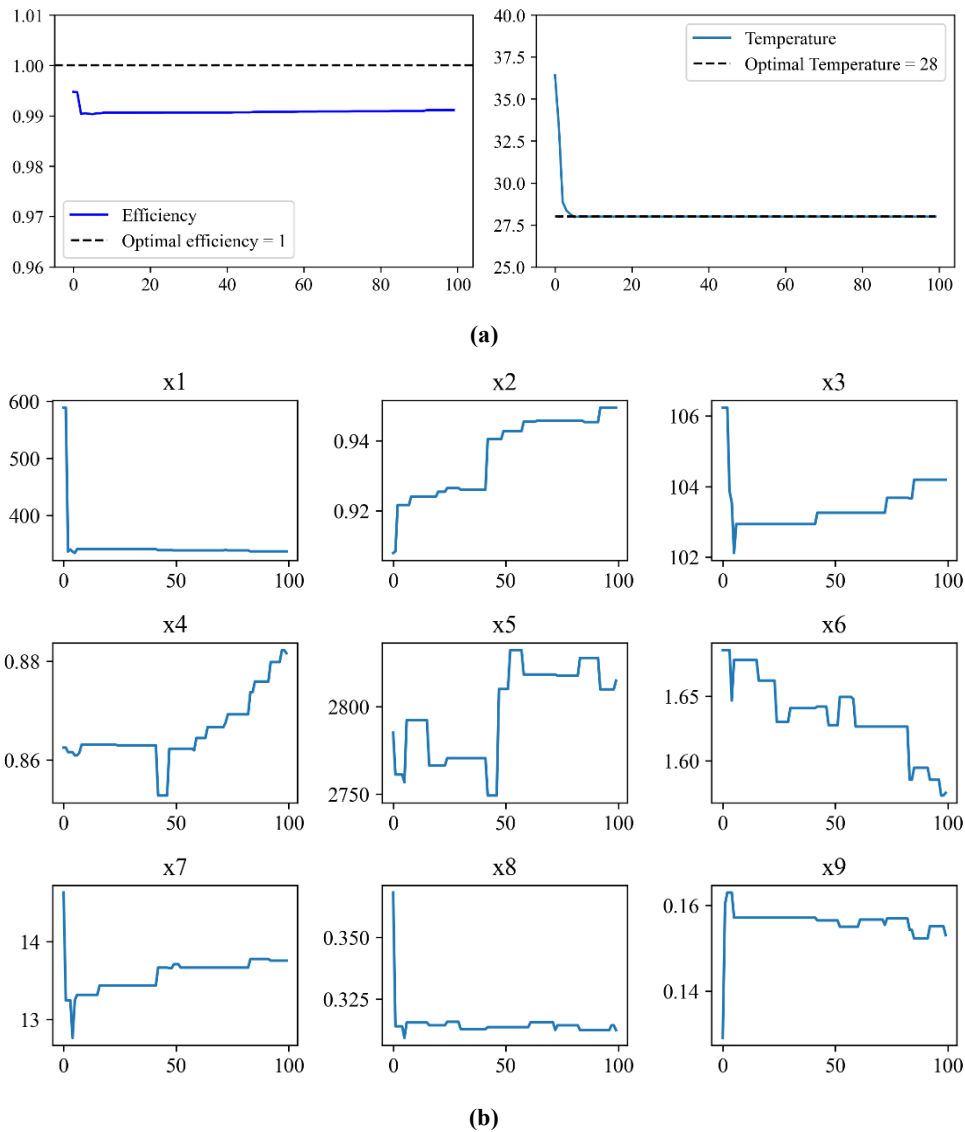

**Figure 10.** Evolution of GA algorithm for (a) optimal efficiency and temperature, and (b) design parameters

### 4.5. Discussion and future work

In this section, we will perform a detailed analysis of each of the component in our proposed surrogate model namely classification of design, probabilistic prediction of efficiency and temperature, and optimization of the design. Specifically, we will discuss the results presented and will support them with additional evidence. Finally, we will discuss the final design of the surrogate model itself with directions on the future developments.

### 4.5.1. Classification of design using classic and neural network-based methods

As described in the previous sections, the goal of classifying each design instance is to verify its practicality. Various algorithms such as logistic regression, decision trees, random forests, XGBoost, and ANN are compared, and various metrics were reported.

Overall results analysis indicated that ANN is superior to the rest of the methods. Results from Table 2 show that ANN outperformed rest of the algorithm across different validation sets on all five folds during cross validation with accuracy of 99±0.3% and binary cross-entropy of 0.035±0.003. This indicates that ANN demonstrated superior generalization and robustness highly robust to the varying validating set compared to its close competitor, XGBoost. XGBoost have similar variance in error and accuracy across the folds but ANN boosts accuracy by average of 2%. For our application, a higher False Positive rate is not desired because it will lead to feasible designs being classified as not feasible, resulting in sub-optimal design. It turns out that ANN outperformed XGBoost in classifying the extremely ambiguous cases of design with False Positive rate = 21 compared to XGBoost's False Positive rate = 58. Table 3 shows that ANN model outperformed all other classification algorithms across every metric, achieving the highest accuracy (0.995), F1 score (0.995), precision (0.994), recall (0.997), and AUC-PR (0.999), while also recording the lowest binary cross-entropy (BCE = 0.029). This indicates not only superior classification performance but also excellent confidence calibration, making ANN the most robust and reliable model for the task. In addition to the metrics, the learning curve for ANN in Figure 6 indicates a stable learning process and consistent improvement in the loss. Therefore, ANN is the best suited model for classifying the design parameters. ANN will be used as the classifier in the objective function for optimization defined in Algorithm 1 (section 3.4.1).

### 4.5.2. Probabilistic prediction of efficiency and temperature

Along with pointwise prediction, a probabilistic prediction algorithm is concerned with modeling the underlying uncertainty by approximating data distribution. The quantification of uncertainty also enables us to investigate confidence in predictions which gives additional insights.

Among the evaluated probabilistic models, the Natural Gradient Boosting with Extra Trees (NGB-ET) demonstrated the most balanced and robust performance in both pointwise prediction accuracy and uncertainty quantification. For efficiency prediction, NGB-ET achieved a RMSE of 0.02 and an R² score of 1.00, indicating near-perfect prediction accuracy and complete variance explanation. Additionally, it maintained a narrow MPIW of 0.04 while achieving a PICP of 0.95, suggesting well-calibrated uncertainty intervals that confidently capture true values. CRPS of 0.01 further confirms the high quality of its probabilistic forecasts. In temperature prediction, although the RMSE of 4.95 is slightly higher than its closest competitor, NGB-ET maintains a strong R² of 0.97 and achieves a PICP of 0.96, with a moderately wide interval (MPIW = 15.95). These results indicate that NGB-ET provides a favorable balance between interval sharpness and reliability, making it a strong candidate for high-stakes applications requiring both accuracy and uncertainty awareness.

The calibration curves in Figures 11(a) and 11(b) provide additional evidence of the strong probabilistic performance of the NGB-ET model. In Figure 11(a), corresponding to efficiency prediction, the actual coverage closely aligns with the ideal calibration line, indicating that the model's prediction intervals are well-calibrated across varying confidence levels. Figure 11(b), for temperature, also shows good alignment at high confidence levels (≥90%), with minor under coverage at intermediate levels. These trends are consistent with the high PICP values reported in Table 4 and confirm that NGB-ET produces not only accurate point predictions but also reliable uncertainty estimates, which are essential for robust surrogate modeling.

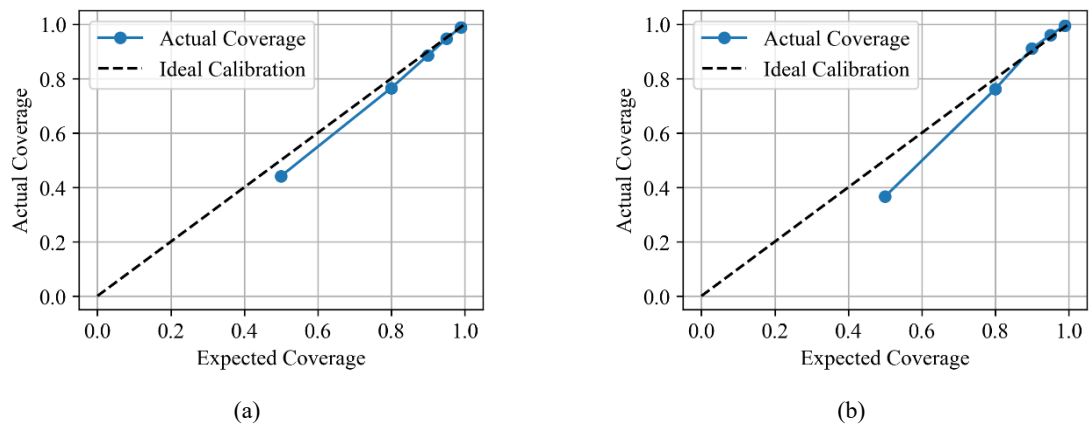

**Figure 11.** Calibration curve for Natural Gradient Boosting predictions for (a) efficiency and (b) temperature

The Histogram of Interval Width (HIW) plots in Figures 12(a) and 12(b) further support the effectiveness of the NGB-ET model in providing sharp and informative prediction intervals. In Figure 12(a), corresponding to efficiency, the majority of interval widths are concentrated around a very narrow range (mostly below 0.002), indicating highly confident and precise predictions. In Figure 12(b), for temperature, the interval widths are also tightly distributed, primarily between 10 and 30, with very few extreme outliers. This reflects a strong balance between sharpness and coverage, consistent with the low MPIW values reported in Table 4. Together, these distributions confirm that NGB-ET produces compact yet reliable uncertainty intervals, which is critical for surrogate modeling tasks that require both prediction confidence and model efficiency.

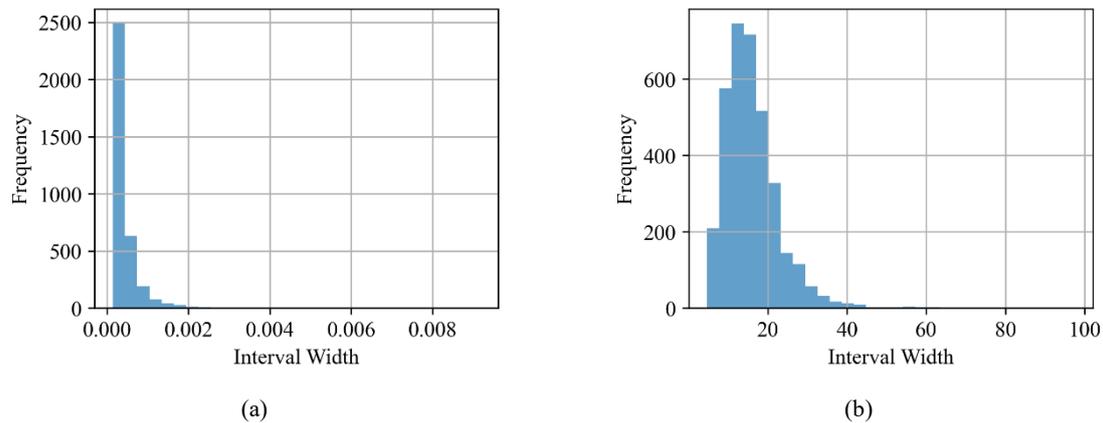

**Figure 12.** Histogram of Interval Widths (HIW) for NGB-ET model predictions for (a) Efficiency and (b) Temperature

The combined results from evaluation metrics, calibration curves, and interval width distributions clearly demonstrate that the NGB-ET model offers a well-rounded probabilistic performance. It achieves high pointwise accuracy, produces well-calibrated prediction intervals, and maintains sharp uncertainty bounds across both efficiency and temperature targets. These strengths make NGB-ET a highly suitable choice for surrogate modeling tasks where both predictive reliability and uncertainty quantification are critical.

### 4.5.3. Optimization of design parameters to obtain optimal performance metrics

PEC design parameter optimization, we do not focus just on fitness minimization but also on complexity and reliability. Accuracy refers to obtaining valid performance metrics while reliability refers to obtaining consistent optimal performance metrics estimate regardless of the design parameter values.

Based on the optimization results in Table 5, the Genetic Algorithm (GA) stands out as the best-performing model. It achieved the lowest fitness value of 0.00004, which is the principal metric to be minimized under the defined multi-objective fitness function. This low score reflects a highly effective trade-off between maximizing efficiency and minimizing deviation from the target temperature, while also satisfying feasibility constraints encoded through classification confidence and prediction intervals. The GA reached a final solution characterized by 99.11% efficiency and a temperature of 28.00 °C, indicating that it not only precisely satisfied thermal constraints but also produced a high-quality, energy-efficient design. When compared to other algorithms, the Genetic Algorithm maintained a clear advantage in terms of fitness minimization. Particle Swarm Optimization (PSO), while delivering a competitive outcome of 99.04% efficiency and 28.01 °C temperature, resulted in a slightly higher fitness score of 0.00019, suggesting that its solution was marginally less aligned with the optimization objective. Stochastic Hill Climbing (SHC) yielded the highest efficiency of 99.16% and the lowest temperature (27.75 °C), but its fitness value of 0.02523 was significantly higher than that of GA. This indicates that while SHC excelled in raw performance metrics, the uncertainty and feasibility penalties embedded in the fitness function reduced its overall optimization quality.

Other algorithms, such as Simulated Annealing (SA) and Tabu Search (TS), performed notably worse. SA reached a moderate efficiency of 99.12% with a higher temperature of 29.14 °C, but its fitness was substantially higher at 0.52636. Tabu Search, despite achieving 99.56% efficiency, produced the worst temperature (31.34 °C) and an extremely poor fitness value of 4.49218, indicating failure to balance objectives within acceptable bounds. In conclusion, the Genetic Algorithm demonstrated the most favorable convergence toward the defined objective, outperforming all other methods in terms of the unified fitness score. Its ability to maintain strong efficiency while tightly controlling temperature and adhering to feasibility constraints affirms its position as the most optimal and reliable algorithm for the PEC design problem under the current surrogate-assisted optimization framework.

### 4.5.4. Future work

While the proposed surrogate modeling framework demonstrates strong performance in PEC design optimization, several improvements are possible. Integrating hardware-in-the-loop (HIL) simulations or experimental validation using physical prototypes would help evaluate the model under real-world conditions, including electromagnetic interference and thermal stress. Expanding the framework to support more complex converter topologies (e.g., full-bridge, multi-level) would improve versatility, though it may require tailored features, broader datasets, and topology-specific constraints. Likewise, incorporating additional performance objectives such as switching losses, voltage ripples, electromagnetic compatibility, or cost could enable more holistic multi-objective optimization. On the algorithmic side, embedding physics-informed machine learning and active learning could improve interpretability and sample efficiency, especially in high-uncertainty regions. Finally, efforts toward model compression, runtime optimization, and deployment on embedded systems are essential for real-time industrial applications.

## 5. Conclusions

This work presents a novel probabilistic surrogate modeling framework that integrates classification, regression, and stochastic optimization to support efficient and reliable design of power electronic converter (PEC) parameters. The proposed system was validated using a simulated half-bridge converter under diverse operating conditions, generating a rich dataset of 30,000 instances across nine key design variables. Empirically, the classification component—implemented using an artificial neural network—achieved outstanding accuracy (up to 99.5%) and minimal binary cross-entropy loss, significantly outperforming traditional methods such as decision trees and logistic regression. For quantity prediction, Natural Gradient Boosting (NGB) demonstrated the best overall performance, achieving near-zero RMSE for efficiency (0.00) and competitive predictive accuracy for temperature (RMSE of 8.44), all while maintaining robust uncertainty calibration (PICP of 1.00). The optimization phase compared several stochastic algorithms, including Genetic Algorithm (GA), Particle Swarm Optimization (PSO), and Stochastic Hill Climbing (STC). While GA achieved the lowest fitness value (0.00013) with 98.61% efficiency and 27.99 °C temperature, STC achieved the highest efficiency (99.16%) and lowest temperature (27.75 °C) with a significantly shorter runtime (1.82 minutes vs. 38 minutes for GA), demonstrating its advantage in convergence speed and computational efficiency. These empirical findings confirm the framework's ability to deliver accurate, uncertainty-aware predictions while efficiently navigating the complex design space of PECs. This makes the proposed approach not only theoretically sound but also practically valuable. Future work will aim to extend this surrogate modeling pipeline to real-time hardware validation, integrate additional physical constraints, and explore generalization across different converter topologies.